\documentclass[aps, twocolumn, prl, showpacs,preprintnumbers,amsmath,amssymb,superscriptaddress]{revtex4-1}
\usepackage{amsmath,amsfonts,amssymb,graphics,graphicx,epsfig,color,times,indentfirst,layout,hyperref,subfigure,multirow,bm}
\usepackage{hyperref}
\usepackage{bbold}
\usepackage{ulem}
\hypersetup{linktocpage,,colorlinks=true,citecolor=red}

\usepackage{soul}

\newcommand{\be}{\begin{equation}}
\newcommand{\een}{\end{equation*}}
\newcommand{\bs}{\begin{split}}
\newcommand{\ben}{\begin{equation*}}
\newcommand{\ee}{\end{equation}}
\newcommand{\es}{\end{split}}
\newcommand{\bmx}{\begin{array}}
\newcommand{\emx}{\end{array}}
\newcommand{\bea}{\begin{eqnarray}}
\newcommand{\bean}{\begin{eqnarray*}}
\newcommand{\eea}{\end{eqnarray}}
\newcommand{\eean}{\end{eqnarray*}}
\newcommand{\dg}{^{\dagger}}
\newcommand{\dn}{^{\vphantom{\dagger}}}

\newcommand{\bp}{{\bf p}}

\newcommand{\ua}{\uparrow}
\newcommand{\da}{\downarrow}
\newcommand{\bb}[1]{\mathbb{#1}}

\newcommand{\eps}{\epsilon}

\newcommand{\sgn}[1]{{\rm sign}{#1}}
\newcommand{\pref}[1]{(\ref{#1})}

\newcommand{\intinf}[1]{\int_{-\infty}^{+\infty}{#1}}
\newcommand{\intoinf}[1]{\int_{0}^{\infty}{#1}}

\newcommand{\intob}[1]{\int_{0}^{\beta}{#1}}

\newcommand{\im}[1]{{\rm Im}\left[ #1 \right]}
\newcommand{\tr}[1]{{\rm Tr}\Big[ #1 \Big]}

\newcommand{\braket}[1]{\left\langle #1\right\rangle}

\newcommand{\mat}[1]{\left(\bmx{cc}#1\emx\right)}

\newcommand{\bw}[1]{\begin{widetext}}
\newcommand{\ew}[1]{\end{widetext}}

\newcommand{\red}[1]{{\color{red} #1}}
\newcommand{\rem}[1]{}
\newcommand{\gray}[1]{}

\begin{document}
\title{Emergent critical charge fluctuations at the Kondo break-down of Heavy Fermions}
\author{Yashar Komijani}
\affiliation{Department of Physics and Astronomy, Rutgers University, Piscataway, New Jersey, 08854, USA}

\author{Piers Coleman}
\affiliation{Department of Physics and Astronomy, Rutgers University, Piscataway, New Jersey, 08854, USA}
\affiliation{Department of Physics, Royal Holloway, University of
London, Egham, Surrey TW20 0EX, UK}
\date{\today}
\begin{abstract}
One of the challenges in strongly correlated electron systems, is to
understand the anomalous electronic behavior that develops at an
antiferromagnetic quantum critical point (QCP), a phenomenon that has
been extensively studied in heavy fermion materials. Current theories
have focused on the critical spin fluctuations and associated
break-down of the Kondo effect.  Here we argue that the abrupt change
in Fermi surface volume that accompanies heavy fermion criticality
leads to critical charge fluctuations. Using a model one dimensional
Kondo lattice in which each moment is connected to a separate
conduction bath, we show a Kondo breakdown transition develops between
a heavy Fermi liquid and a gapped spin liquid via a QCP with
$\omega/T$ scaling, which features a critical charge mode directly
associated with the break-up of Kondo singlets. We discuss the
possible implications of this emergent charge mode for experiment.
\end{abstract}
\maketitle

{\it Introduction} - The relation between valence
fluctuations and the Kondo effect
has long fascinated the physics community\,\cite{Schrieffer66}.  A partially occupied 
atomic state, weakly hybridized with a conduction sea,
forms a local moment\,\cite{Anderson78} and its virtual valence fluctuations give rise to 
low frequency spin-fluctuations, 
while leaving its charge essentially frozen. On the other hand,
in heavy fermion systems, the Kondo-screening of the local moments
gives rise to an enlargement of
the Fermi surface, a phenomenon that is well 
established both theoretically\,\cite{Yamanaka97,Oshikawa00} and through
Hall coefficient\,\cite{Paschen04},
quantum oscillation\,\cite{Shishido05}, angle-resolved photoemission spectroscopy, and scanning tunnelling microscopy 
measurements\,\cite{qiuyun17,Aynajian2012}.
The large Fermi surface of a Kondo lattice is believed to partially collapse
when Kondo screening is
disrupted\,\cite{Coleman:2001hc,Si2001,Senthil2003,
Senthil2004,Coleman05m,Paul07,Pepin07,Nejati17}
at an antiferromagnetic (AFM)  quantum
critical point (QCP), a phenomenon
known as  ``Kondo breakdown'' (KBD). 

 Recently, a number of experiments have observed a coincidence of
critical charge fluctuations at the magnetic quantum critical points
in CeRhIn$_5$\,\cite{Ren17} YbRh$_2$Si$_2$\,\cite{Prochaska18} and
$\beta$-YbAlB$_4$\,\cite{Kobayashi18}. 
\begin{figure}[tp!]  \includegraphics[width=1\linewidth]{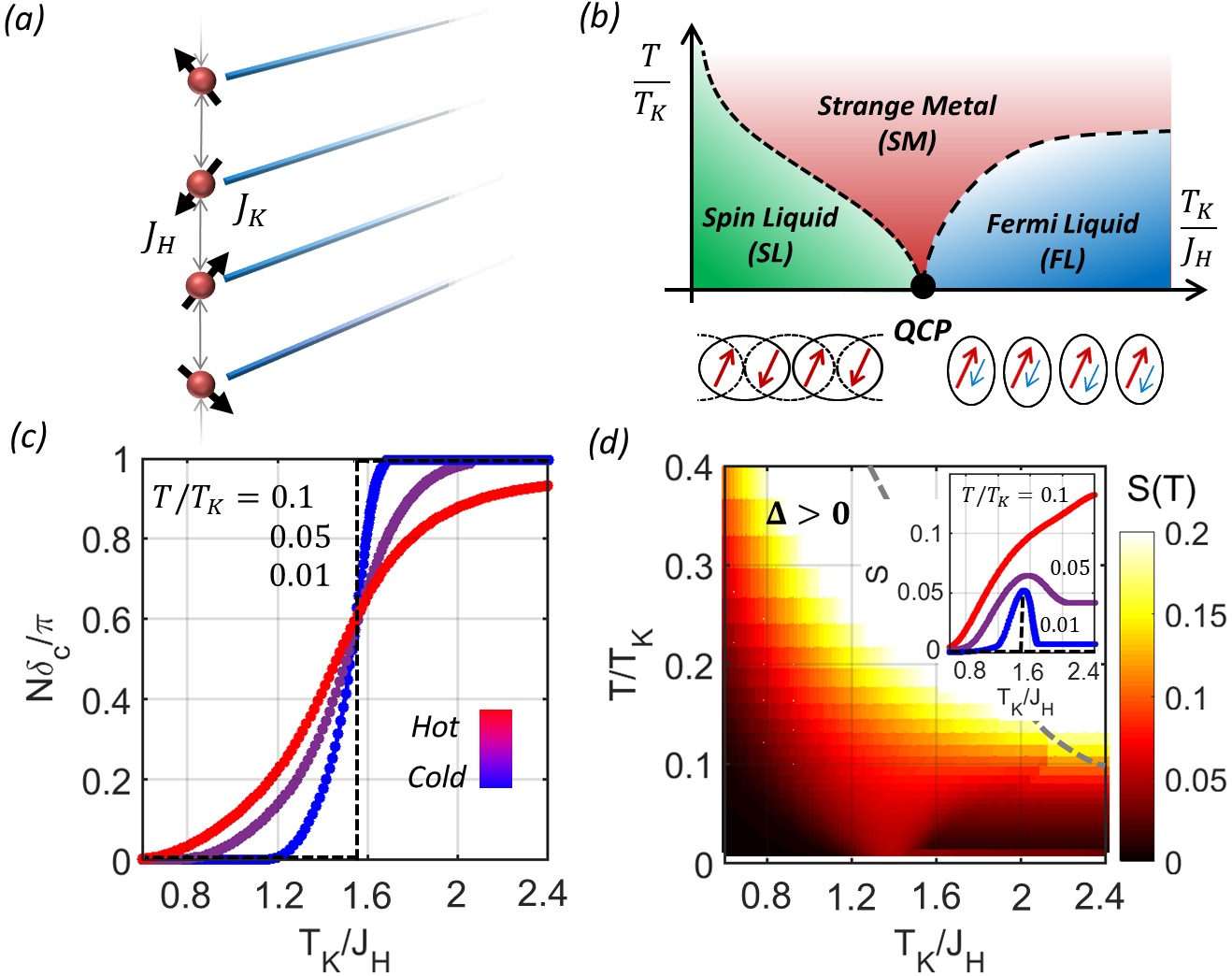}
\caption{\small (color online) (a) Model 1D Kondo lattice, 
with local moments (red) with an AFM Heisenberg 
coupling $J_H$, individually screened by 
separate conduction electron baths (blue wires). (b) 
Schematic phase diagram showing the transition between heavy-Fermi
liquid (FL) and spin-liquid (SL) phases at a QCP {which evolves to a fan of strange metal (SM) at finite temperature}. 
(c) Conduction electron phase shift
$\delta_c$ as a function of $T_K/J_H$, which extrapolates to a
step-like jump from $0$ to $\pi/N$ as $T\rightarrow 0$.
(d) Color map of entropy $S (T)$,
showing the collapse of energy scales at the QCP:
the dashed line separates localized ($\Delta =0$ to the right) and delocalized
($\Delta >0$ to the left) spinon regimes. Inset:
temperature cuts showing the accumulation of entropy
at the QCP. }\label{fig1}
\end{figure}
Watanabe and Miyake have argued that the development of soft charge fluctuations near a heavy fermion QCP is likely a result of a
quantum-critical end-point, in which a first-order valence 
changing transition line is suppressed to low temperatures
\cite{Watanabe2013,Miyake2014,Watanabe14,Watanabe2015,Scheerer:2018du}.
Here we present an alternative view, arguing that the coincidence of soft charge fluctuations and
Kondo breakdown is a natural consequence of the Fermi surface collapse.

In the eighties, Anderson introduced the concept of 
a {\it nominal valence} to 
distinguish the valence of a rare earth ion
infered from the apparent delocalization of f-electrons
\cite{Anderson83,revfiskaeppli}, from the
core-level valence, infered from spectroscopy. 
From this perspective, 
a shift in nominal valence is
associated with formation of a large Fermi surface, even in a
strict Kondo lattice where the core-level valence is fixed. 
Interpreted literally, this implies a kind 
of many-body {\it ionization} in the Kondo lattice, 
in which a fractionalization of  local moments into charged heavy electrons,
leaves behind a 
compensating positive background of Kondo singlets
\cite{LEBANON20081194}. 
Taken to its logical extreme, such an interpretation would then imply 
that at KBD quantum critical point, 
degenerate fluctuations in the nominal valence 
will give rise to an observable soft charge mode.

While Kondo Breakdown 
has been extensively 
modelled at an impurity-level\,\cite{Pixley2012,Chowdhury15}
and simulated using dynamical mean-field
theory\,\cite{Si2001,Leo08,Martin10}, a possible link with charge
fluctuations has not sofar been explored {in the lattice}. 
To examine this idea, we introduce a simple field-theoretic framework
for Kondo breakdown, emplying a Schwinger boson
representation of spins that permits us to treat 
Kondo screening and antiferromagnetism
\cite{Parcollet1997,Arovas1988}. Early application of this method 
demonstrated its efficacy
for describing a ferromagnetic quantum critical
point\,\cite{Komijani18} in a Kondo lattice. 
Here we consider a Kondo screened one dimensional (1D) AFM
[Fig.\,\ref{fig1}a], examining the quantum phase transition
transition between a spin-liquid and a Fermi liquid\,[Fig.\,\ref{fig1}(b)]. The conduction electron phase shift
(related to the Fermi surface size) jumps at $T=0$ [Fig.\,\ref{fig1}(c)], indicating that
QCP is a KBD transition. Additionally, we find that the KBD features a zero point entropy [Fig.\,\ref{fig1}(d)].
In our calculations we observe that the 
KBD is linked to the emergence of a gapless charge degree
of freedom at the QCP which occurs in natural coincidence with a
divergent charge and staggered spin susceptibility. 

{\it Model -} The simplified 1D Kondo lattice 
is a chain of antiferromagnetically coupled spins 
each individually screened by a conduction electron bath:
\begin{equation}\label{}
H = \sum_j
\Big[ H_{C} (j) + J_{K} \vec{S}_{j}\cdot \vec{\sigma }_{j} + J_{H}\vec{S}_{j}\cdot\vec{S}_{j+1}\Big].
\end{equation}
Here  $\vec{S}_{j}$ is the spin at the $j$-th site, 
coupled antiferromagnetically to its neigbor with strength $J_{H}$.
$H_C (j) =\sum_{{\bf p}}\eps_{\bp } c\dg_{\bp\alpha} (j)
c_{\bp \alpha} (j)$ describes the conduction bath coupled to the
$j$-th moment  in the chain, where $\bp $ is the momentum of the
conduction electron. 
$\vec\sigma_{j} = \psi \dg_{j\alpha } \vec{\sigma }_{\alpha \beta
}\psi_{j\beta } $ is the spin density at site $j$, where
$\psi\dg _{j\alpha } = 
\sum_{ \bp }c\dg _{\bp \alpha } (j)$ creates an electron on the chain
at site $j$. 

{\it Global phase diagram} -  Numerical and
experimental studies of heavy-fermion systems are often interpreted \cite{Si2001,coleman_jphys,Coleman2010} 
within  a 
global phase diagram of the Kondo lattice, with two axes: a
Doniach parameter $x=T_K/J_H$ \cite{Doniach1977}, where $T_{K}$ is the
Kondo temperature, and a frustration parameter
$y$ representing the magnitude of quantum fluctuations, controlled by
geometrical or dimensional frustration.  The 1D limit
provides a way to explore the  two extremes of $y$:
on the one hand, the uniform magnetization of a 1D FM
commutes with the Hamiltonian and has no quantum
fluctuations, corresponding to $y=0$ \cite{Komijani18}, whereas a 1D AFM never
develops long range order, loosely corresponding to $y=\infty $.
{When the magnetic coupling is Ising-like, both models can be
mapped to the dissipative transverse-field Ising model.  But a
Heisenberg magnetic coupling has been proven to be difficult to treat
with these methods\,\cite{Lobos2012,Lobos2013} and a single formalism that can access various phases and critical points is highly desirable.}


{\it The method} - We use a  large-$N$ approach, obtained by
enlarging the spin rotation group from SU(2) to SP($N$),
representing the spin $S$ local
moments using Schwinger bosons (``spinons''), according to 
$S_{\alpha\beta}=b\dg_\alpha
b\dn_\beta-\tilde\alpha\tilde\beta
b\dg_{-\beta}b\dn_{-\alpha}$\cite{Read91,Flint2009hr}, where
$\alpha\in[\pm 1, \dots \pm  N/2]$, $\tilde\alpha=\sgn(\alpha)$ and
$n_{b} (j)=2S$ is the number of bosons per site. Each moment is 
coupled to a $K$-channel conduction sea, with Hamiltonian
\begin{equation}\label{}
H=\sum_j\Big[
H_{AFM} (j) +H_{K} (j) +H_{C} (j)+H_{\lambda}(j)\Big], 
\end{equation}
where 
\bea
H_{AFM} (j)&=&- (J_H/N)(\tilde\alpha b\dg_{j\alpha}b\dg_{j+1,-\alpha})(\tilde\beta b\dn_{j+1,-\beta}b\dn_{j\beta}),
\nonumber\\
H_K (j)&=&-(J_K/N)\bigl (b\dg_{j\alpha}\psi \dn_{ja\alpha} \bigr )
(\psi\dg_{ja\beta}b\dn_{j\beta}),
\nonumber\\
H_{\lambda}(j)&=&\lambda_{j} [n_{b} (j)-2S]\label{eq3}.
\eea
Here we have adopted a summation convention for the repeated 
greek $\alpha\in[\pm 1,\pm N/2]$ 
spin and roman $a\in [1,K]$ channel
indices. The Lagrange multiplier $\lambda_{j}$ imposes the 
constraint $n_{b} (j)=2S$: we take {$2S=K=2sN$} for perfect screening,
where $s$ is kept fixed. 

\begin{figure}[tp] \includegraphics[width=1\linewidth]{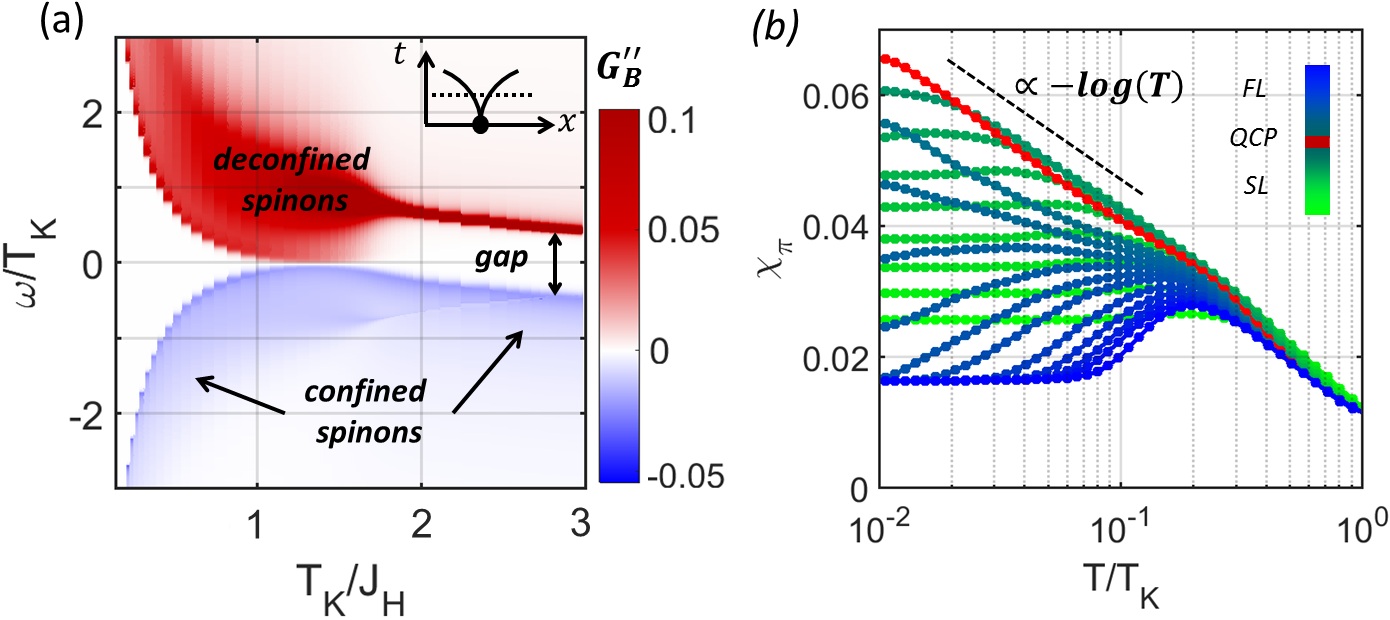}
\caption{\small (color online) 
(a) Calculated spectral function of
spinons $G_B''(\omega-i\eta)$ at $T/T_K=0.03$ shows confined spinons protected by a gap in the FL
and SL, and deconfined with with a soft excitation 
gap in the SM regime. (b) Staggered spin susceptibilities vs. $T/T_K$
for various various values of $T_K/J_H$ from SL (in green)
and FL (in blue) passing QCP (in red). A $\log$-divergence at the QCP is visible.  }\label{fig2}
\end{figure}

We carry out the Hubbard-Stratonovich
transformations:
\bea
H_K (j)&\to&
\bigl [
(b\dg_{j\alpha}
 \psi \dn_{ja\alpha})
\chi_{ja} 
+{\rm h.c}\bigr ]+\frac{N\bar\chi_{ja}\chi\dn_{ja}}{J_K}\\
H_{AFM} (j)&\to&
\bigl [\bar  \Delta_{j}(\tilde\alpha b\dg_{j+1,-\alpha} b\dg_{j,\alpha})+{\rm h.c}\bigr ]
+\frac{N  |\Delta_{j}|^{2}}{J_H}\nonumber,\label{eq4}
\eea
where $\chi_{ja}$ is  a Grassmanian ``holon''  field that mediates the Kondo effect
at site $j$ in channel $a$, while $\Delta_{j}$ describes the
development of singlets between site $j$ and $j+1$. {See \cite{Komijani18} For a discussion of spurious 1st order transition and their remedy.}

A mean-field  resonating valence-bond (RVB) description of the 1D magnetism is obtained
from a uniform mean-field theory where 
$\Delta_{j}=i\Delta_B/2$, 
and $\lambda_{j}=\lambda
$, giving rise to a bare spinon dispersion
$\eps_{B} (p) =[{\lambda^2-\Delta_p^2}]^{\frac{1}{2}}$, with 
$\Delta_p=\Delta_B\sin p$. 
Both $b$ and $\chi$ fields have non-trivial dynamics\,\cite{Parcollet1997,Parcollet1998,
Coleman2005,Rech2006,Komijani18}, with self-energies 
\be
\hspace{-.25cm}\Sigma_\chi(\tau)=g_0(-\tau)G_B(\tau),\qquad \Sigma_B(\tau)=-\gamma g_0(\tau)G_\chi(\tau).
\ee
Here, $\gamma=K/N=2s$ and $G_{\chi } (\tau )$, $G_{B} (\tau )$ and $g (\tau )$ are the
local propagators of the holons, spinons and conduction electrons,
respectively. 
The conduction electron self-energy
is 
of order
${\cal O}(1/N)$ and is neglected in the large-$N$ limit, so that
$g_{0}(\tau)$ is the bare local conduction
electron propagator.
The holon Green's function $G_\chi (z)=[{-J{_K}^{-1}-\Sigma_\chi(z)}]^{-1}$,
is purely local, whereas the spinons are delocalized by the RVB pairing
with propagator 
$\bb G_{B} (p,z)= [z\tau^{z}-\lambda\bb 1-\Delta_p\tau^{{x}}-\bb\Sigma_{B} (z)]^{-1}$. 
The self-energy $\bb\Sigma_{B} (z)$ is diagonal in Nambu space, and the momentum sum in
$\bb G_{B}^{\rm loc}(z)=\sum_p \bb G_B(p,z)$ can be done analytically\,\cite{SM}.


Stationarity of the free energy with respect to $\lambda$ enforces the
mean-field constraint $\langle  n_{b} (j)\rangle  =K$,  
and with respect to $\Delta_B$ determines the
relation $\Delta_B(J_H)$\,\cite{SM}. We solve these self-consistent
equations numerically on the real-frequency axis using linear and logarithmic grids.

{\it The two limits} - In absence of Kondo screening (when
$T_K/J_H$ is small) the constraint is satisfied with $\lambda>\Delta_B$. This Schwinger boson model describes a bipartite
spin chain, in which each sublattice is in the symmetric
spin-$S$ representation of SP($N$)\,\cite{Read91,Flint2009hr}: Each spin can form
singlets with its neighbors in an RVB state for any value $S$. This, together with the Gutzwiller
projection treated by a soft constraint leads to a U(1) gapped spin
liquid\,\cite{Arovas1988}, closely analogous to the integer-spin
Haldane chain\,\cite{Haldane83}. 
A lattice with
closed boundary condition has a unique ground state and corresponds to
a symmetry protected topological phase\,\cite{Pollmann10}.

The large $T_K/J_H$ limit corresponds to a local Fermi
liquid\,\cite{Rech2006,Komijani18} at each site of the chain, in which
the electrons and spinons
form bound, localized singlets, protected by a
spectral gap of the size $T_K$; the remaining electrons are
scattered with a phase shift $\delta_{c}=\pi/N$. Fig.\,(\ref{fig1}b)
summarizes the phase diagram as $T_K/J_H$ is varied
between the above two limits, which we discuss in the following.

{\it Ward identity, Entropy, Phase Shifts -} At large $N$, the many
body equations can be derived from a Luttinger Ward functional,
leading to an exact relation between the conduction electron and holon
phase shifts $\delta_c=\delta_\chi/N$ and a closed form formula for
the entropy\,\cite{Coleman2005,Rech2006}.
Fig.\,\ref{fig1}(c) shows the conduction electron phase shift
$N\delta_c/\pi$ as a function of
$T_K/J_H$. In the Fermi liquid, $\delta_\chi=N\delta_c$ is equal to $\pi$,
equivalent to a large Fermi surface, but it is zero
in the spin liquid regime. Extrapolating the calculations to $T\to 0$, the phase shift appears to jump at the QCP separating
the spin-liquid (decoupled electrons) and the Fermi-liquid.
From the perspective of conduction electrons, both SL and FL phases are Fermi liquids and the transition in $\delta_\chi$ is a measure of change in the Fermi surface, a manifestation
of Kondo breakdown (KBD).

{\it Entropy - } Fig.\,\ref{fig1}(d) shows the colormap of the 
entropy $S (T)$ across the phase diagram. The gray dashed line
indicates a second order phase transition for the internal variable
$\Delta_B$ that separates a local Fermi liquid ($\Delta_B=0$) from a
de-localized regime ($\Delta_B>0$). The collapse of the energy scale
from both sides are visible.
Unlike the 1D ferromagnetic QCP \cite{Komijani18}, 
the antiferromagnetic QCP develops a residual entropy
$S_{E}/N\approx 1/20$ at a spin of $s=0.1$ per moment (inset of
Fig.\,\ref{fig1}d),

{\it Magnetic excitations} - Fig.\,(\ref{fig2}a) shows the spinon
spectrum $G''_B(\omega-i\delta)$ vs. $T_K/J_H$ at $T/T_K=0.03$. Approaching the transition from the Fermi
liquid side (right), the spinon
spectrum shifts to positive frequencies and, maintaining overall gap
size, brings the gap edge close to the chemical potential and only then, the hard
gap closes at the QCP. Passing through the critical point, 
the gap re-opens due to development of 
short-range RVBs in the spin liquid regime. 
Fig.\,\ref{fig2}(b) shows
the temperature dependence of the 
staggered spin susceptibility $\chi_{\pi}$, 
which acquires a
logarithmic temperature dependence 
$\chi_{\pi}\sim -\log T$ at the QCP. The Fermi liquid (blue) exhibits a crossover
from Curie law $1/T$ to a Pauli form $1/T_K$, with a characteristic
peak at $T/T_K\sim 0.1$. As $T_K/J_H$ is reduced the peak position is
unchanged (unlike the 1D FM case\,\cite{Komijani18}) whereas the low temperature
susceptibility develops a logarithmic divergence. Similar divergence is observed in local spin susceptibility but the uniform susceptibility is merely suppressed by the magnetism\,\cite{SM}.

\begin{figure}[t!]  \includegraphics[width=\linewidth]{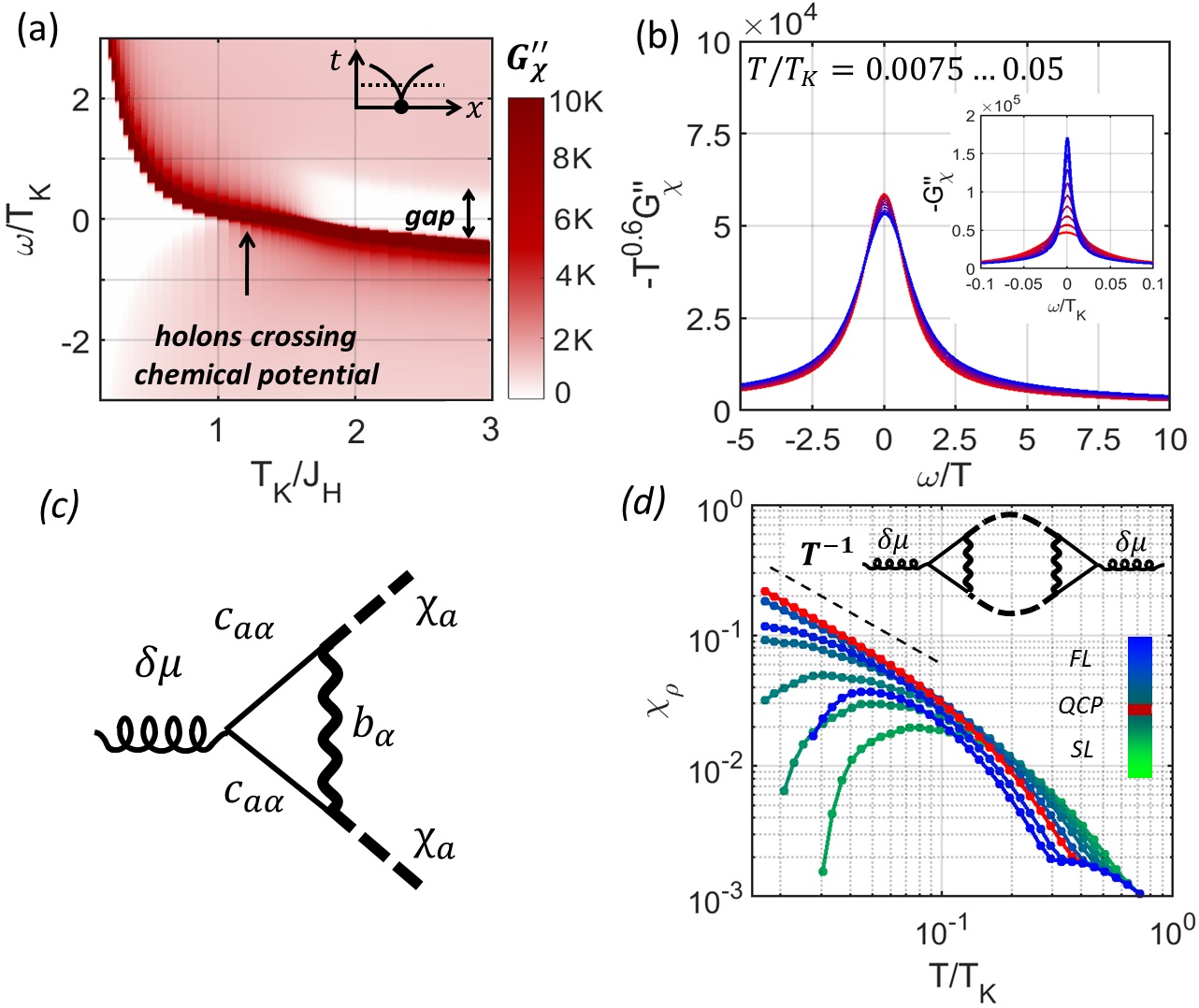}
\caption{\small 
(a) Calculated spectrum of
holons $G_\chi''(\omega-i\eta)$ showing that the holon mode crosses the chemical potential at the QCP. (b) $\omega/T$ scaling of the holon Green's function
$G''_\chi(\omega-i\eta)$ at the QCP. The inset shows the holon mode before scaling. (c) The ${\cal O}(1)$ charge vertex of holons coupling them to potential fluctuations.
(d) The charge susceptibility computed via
this vertex corrections (inset) shows a $T^{-1}$
divergence at the QCP point (red) and its suppression in SL/FL sides.}\label{fig3}
\end{figure}

{\it The holon spectrum} - $G''_\chi(\omega-i\eta)$
shows a striking behavior at the QCP (Fig.\,\ref{fig3}a). Most of the spectral
weight is contained in a sharp holon mode which crosses the chemical
potential as $T_K/J_H$ is tuned from Fermi liquid (right) to
spin-liquid (left). In the critical regime 
at a finite temperature, 
the holon mode is pinned to the Fermi energy over a finite range of
Doniach parameter, which shrinks to a point as $T\to 0$,
forming a strange metal {(SM)} regime at finite temperature 
with deconfined critical holon and spinon modes. 

{\it $\omega/T$ scaling} - At the QCP, the holon mode lies at zero
energy.  Fig.\,\ref{fig3}(b) shows that the holon spectra at different
temperatures collapse onto a single scaling curve
$G''_\chi(\omega,T)=T^{-\alpha}f(\omega/T)$. For $s=0.1$ we find
$\alpha=0.6$ consistent with a scaling analysis\,\cite{SM}. The
universality class of the QCP appears to be that of an overscreened
impurity model\cite{Parcollet1997} , with an effective number of channels
$K_{\rm eff}/N= ( 1/\alpha-1)\approx 0.67>2s$.

\rem{What are the quantum numbers of this critical holon mode and in
particular, does it couple to the electromagnetic field? 
On the one hand, the local moments are neutral and there is 
no direct electromagnetic vertex in the starting model. 
However, as we shall now argue,} 
The holon modes have an emergent
coupling to the electromagnetic field, mediated via the internal
vertices of the lattice Kondo effect.
In particular, the field theory implies an ${\cal
O}(1)$ vertex correction that couples to  the electric potential 
as shown in Fig.\,\ref{fig3}(c). 
At low energies the vertex can be approximated by
$\Gamma_\chi=-\frac{d}{d\omega} \Sigma_\chi$. This quantity perfectly cancels
the wavefunction renormalization of the holon propagator
$G_\chi\approx Z_\chi/ \omega$, where $Z_\chi=-[\partial_
\omega\Sigma_\chi]^{-1}$, so that the holon couples to the
electric potential  with a net charge $\Gamma_\chi
Z_\chi=+1$. Fig.\,\pref{fig3}(d) shows the charge susceptibility calculated using this vertex corrections. At the QCP, the temperature dependence of the
holon charge susceptibilty acquires a Curie-like temperature dependence
$\chi_{\rho}\sim 1/T$.

{\it Discussion} - We have studied a simplified Kondo lattice model in
the large-$N$ limit, enabling us to extract the KBD
physics directly on a lattice. {It is illuminating to note that both specific heat and spin susceptibility \red{\cite{SM}} disagree with the predictions of Hertz-Millis theories of the KBD based on using hybridization as an order parameter \cite{Paul07,Pepin07}.} 

One of the striking features of {our description of the} KBD quantum critical point is the presence of an emergent, spinless critical charge mode
with a Curie-like charge susceptibility. Our model calculations can be extended
in various ways, by going to higher dimensions, by generalizing to the mixed valence regime, and with considerable increase in computation, to a model in which a single
bath is shared between all moments.  In the general Kondo lattice, the
charge conservation Ward identity links the change in the volume of
the conduction electron Fermi surface $\Delta v_{FS}$ to the charge density of the
Kondo singlets, described by the  
holon phase
shift\,\cite{Coleman2005}
\begin{equation}\label{}
  N\frac{\Delta v_{FS}}{(2\pi)^{3}} =
\overbrace {\sum_{\bp}\frac{1}{\pi}{\rm Im}
\ln
  \biggl[-G_{\chi }^{-1} (\bp ,z)\biggr]_{z=0+i\delta}}^{\delta _{\chi }/\pi} 
\end{equation}
Quite generally, the holon phase shift is zero or $\pi$ in the
localized magnetic, or Fermi liquid phases respectively, but 
must jump between these two limits at the quantum critical point {establishing critical holons. This and the O(1) charge vertex leads to critical charge fluctuations, independent of the details of the model.} 
This strongly suggests that the gapless holon mode seen in
our model calculation will persist at a more general 
Kondo breakdown quantum critical fixed point. {Whether the Ward identity remains valid in models with reduced symmetry, is something we leave for future.}

This raises the fascinating question how the predicted critical
charge modes at KBD might be observed experimentally.  One mode of
observation, is via the coupling to nuclear Mossbauer
lines \cite{Komijani2016}. A recent observation of the splitting of the
Mossba\"uer line-shape\, \cite{Kobayashi18}, characteristic of slow
valence fluctuations,  may be a fingerprint of these
slow charge fluctuation.

Another interesting question is whether the residual entropy 
of the QCP might survive beyond the independent bath approximation. 
A residual ground-state
entropy is a signature of infinite-range entanglement, and has been
seen in various quantum models, such as the two channel Kondo
model \cite{Andrei,Tsvelik84,Ludwig94} or the Sachdev-Ye-Kitaev model \cite{Sachdev1992,Georges2001jz,Maldacena16}. 
In the single-channel Kondo problem, the 
Kondo screening length\,\cite{KondoCloud2}
$\xi_K\sim v_F/T^{\rm eff}_K$ plays the role of an entanglement
length-scale, beyond which the singlet ground-state is disentangled from the
conduction sea.  
If the collapse of the Kondo temperature $T_K^{\rm eff}\to 0$ at the QCP
of a Kondo lattice 
involves a divergence of the entanglement length $\xi_K\to\infty$, 
the corresponding quantum critical point would 
be expected to exhibit an extensive entanglement entropy. Such naked quantum
criticality is likely  to be
censored by competing ordered phases that consume the entanglement
entropy of the critical regime, concealing 
QCP beneath a dome of competing phase,
such as superconductivity.   

{Lastly, a peculiar feature of strange metals in heavy-fermions is that the resistivity tends to
be linear in $T$ over a wide range of tempreature. At present, such behavior can be derived only from quenched
disordered models\,\cite{Parcollet99}. One of the fascinating implications of a Curie-law charge
susceptibility $\chi_\rho\sim 1/T$ seen in our calculations, 
is that if combined with a temperature-independent diffusion of incoherent holon motion,  it would give
rise to a Curie conductivity (linear 	resistivity $\rho
\propto 1/\sigma \sim T$) via the Einstein relation $\sigma =D\chi_{c}\sim 1/T$, where $D$ is the holon
diffusion constant. This raises the interesting possibility that
linear resistivities are driven by an emergent critical charge mode. 
}

This work was supported 
by NSF grant DMR-1830707 (Piers Coleman), and by
a Rutgers University Materials Theory postdoctoral fellowhsip
(Yashar Komijani).
We would like to thank S.~Nakatsuji, H.~Kobayashi, A.~Vishwanath,
J.~Pixley, C.~Chung and A.~Georges for stimulating discussions, M.~Oshikawa for helpful correspondence { and A.~M.~Lobos, M.~A.~Cazalilla and P.~Chudzinski for discussing that clarified the relation to \cite{Lobos2012,Lobos2013}.}

\bibliography{AFMKondo}

\section{Supplementary Material}

This supplementary section
contains additional details, proofs and calculations that are not
included in the paper. For completeness, in section A, we include a bosonization mapping between the Ising limit of our model and the dissipative transverse field Ising model. In section B, we review the Arovas-Auerbach
solution to the 1D antiferromagnet using the Schwinger boson
representation of the spin. In C we provide the details of the dynamical large-$N$ equations, as well as formulas used to evaluate various thermodynamical properties. In section D we
present a perturbative analysis of the large-$N$ equations showing
that the phase shift jumps between $0$ and $\pi$.  In section E we
establish that at the QCP the renormalized energies of spinon and
holon go to zero, and show that the critical exponent we found numerically is consistent with the scaling limit of large-$N$ equations. In section F we provide additional data on the holon phase shift
and charge susceptibility to support the statements of the paper. 

\subsection{A. Mapping to the dissipative Transverse-field Ising model}
In this section we use the bosonization technique to show that in the limit of purely Ising coupling between the magnetic moments (both ferromagnetic and antiferromagentic), our model can be mapped to a dissipative transverse-field Ising model. Such a mapping first appeared in \cite{Lobos2012}. This is a property of the independent screening approximation and does not depend on the dimensionality of the spin lattice. The Hamiltonian is
\be
H=-J_H\sum_n \vec S_n\cdot\vec S_{n+1}+J_K\sum_n \vec S_n.\vec s_{n}+H_C.
\ee

Let us write the Hamiltonian in the general anisotropic form
\bea
H&=&-\sum_n\Big[J_H^zS_n^zS_{n+1}^z+{J_H^\perp}\Big(S_n^xS_{n+1}^x+S_n^yS_{n+1}^y\Big)\Big]\\
&&+\sum_n\Big[J_K^zS_n^zs^z_n+\frac{J_K^\perp}{2}\Big(S^+_ns_n^-+h.c.\Big)\Big]+H_C.\qquad
\eea
In the continuum limit, $H_C$ is
\bea
H_C&=&-iv_F\sum_{n\sigma}\intoinf{dx[\psi\dg_{Rn\sigma}\partial_x\psi\dn_{Rn\sigma}-\psi\dg_{Ln\sigma}\partial_x\psi\dn_{Ln\sigma}]}\\
&\to&-iv_F\sum_{n\sigma}\intinf{dx\psi\dg_{Rn\sigma}\partial_x\psi\dn_{Rn\sigma}}.\nonumber
\eea
At $x=0$ we have the boundary condition $\psi_R(x=0)=\psi_L(x=0)$. We have unfolded the semi-infinite wire into a single set of right-movers which interact (via a Kondo interaction) with the impurities at the origin. Next, we bosonize:
\be
\psi_{Rn\sigma}\sim \exp[{i\sqrt{2\pi}\varphi_{Rn\sigma}}],\qquad :\psi\dg_{R}\psi\dn_R:=\frac{\partial_x\varphi_R}{\sqrt{2\pi}}.
 \ee
We can introduce charge and spin bosons $\sqrt{2}\varphi_{c/s}=\varphi_\ua\pm\varphi_\da$ and only spin-bosons appear in the Hamiltonian. The conduction-band spin is given by
\bea
2s^z_n&=&\psi\dg_{Rn\ua}\psi\dn_{Rn\ua}-\psi\dg_{Rn\da}\psi\dn_{Rn\da}\to\frac{\partial_x\varphi_{n,s}}{\sqrt\pi},  \\ s^+_n&=&\psi\dg_{Rn\ua}\psi\dn_{Rn\da}\to\exp[i2\sqrt{\pi}\varphi_{n,s}(0)].
\eea
The bosonized Hamiltonian is then
\bean
H&=&-\sum_n\Big[J_H^zS_n^zS_{n+1}^z+\frac{J_H^\perp}{2}\Big(S_n^+S_{n+1}^-+h.c.\Big)\Big]+H_C\\
&+&\sum_n\Big[\frac{J_K^z}{2\sqrt{\pi}}S_n^z\partial_x\varphi_{n,s}(0)+\frac{J_K^\perp}{2}\Big(S^+_n e^{i2\sqrt\pi\varphi_{n,s}}+h.c.\Big)\Big].
\eean
We also note that
\be
H_C=\frac{v_F}{2}\sum_n\intinf{[(\partial_x\varphi_c)^2+(\partial_x\varphi_s)^2]}.
\ee
Next we apply the Emery-Kievelson transformation to simplify the Kondo interaction
\be
U_\alpha=\prod_n\exp[i\alpha S^z_n\varphi_{n,s}(0)].
\ee
Using the identities
\bea
U\dg \frac{\partial_x\varphi_s}{\sqrt{2\pi}} U&=&\frac{\partial_x\varphi_s}{2\sqrt\pi}-\frac{\alpha}{\sqrt{2\pi}}S^z\delta(x),\\
U\dg S_n^+ e^{i2\sqrt\pi\varphi_s}U&=&S_n^+e^{i(2\sqrt\pi-\alpha)\varphi_s},
\eea
and choosing $\alpha=2\sqrt\pi$, the transverse Kondo coupling is eliminated, and we then have
\bea
U\dg H U&=&-\sum_n J_H^zS_n^zS_{n+1}^z+ \sum_n\Big[J_K^\perp S_n^x\\
&&+(\frac{J^z_K}{2\sqrt{\pi}}-\frac{\alpha}{\sqrt{2\pi}\rho})S^z_n\partial_x\varphi_{n,s}\Big]+H_C+H_{\perp}[J_H^\perp].\nonumber
\eea
We have only explicitly written the Ising part of Ferromagnetic coupling and have moved the remaining terms into $H_{\perp}$. If we choose $\rho J^z_K=\sqrt{2}$, the Ising part of Kondo interaction then drops out and the transverse Kondo interaction appears as a Zeeman field acting on the ``dressed spin''. This is called the ``Decoupling point'' and constitutes an additional solvable point of the single-impurity Kondo model, similar to the ``Toulouse point''. The Kondo temperature of this highly anisotropic Kondo system is $T_K=J_H$. Therefore, assuming $J_H^\perp=0$, we have 
\be
\hspace{-.23cm}J_H^\perp=0, \qquad \to \quad H=-\sum_n J_H^zS_n^zS_{n+1}^z+ \sum_n T_K S_n^x,
\ee
which is the transverse-field Ising model. This model can be re-written using a Jordan-Wigner transformation in terms of two Majorana fermions with a gap that changes sign at the QCP separating the disordered and ordered phases. Moving away from the Decoupling point introduces an additional $\delta S^z\partial_x\varphi$ term into the Hamiltonian which implies that the spins are now dissipatively coupled to a gapless bosonic bath.\\

If $J_H^\perp\neq 0$, we have
\bean
U\dg H_{\perp}U&=&-\frac{J_H^\perp}{2} \sum_n\Big(S_n^+S_{n+1}^- e^{i\alpha[\varphi_{sn}-\varphi_{s,n+1}]}+h.c.\Big).
\eean
 At
the strong Kondo coupling fixed point, Ref.\,\cite{Lobos2012} projects out the
magnetic moments, mapping the problem to a dissipative Josephson
Junction array which is further mapped to a Hertz-Millis theory. This
 corresponds to a spin-density instability of heavy-fermions [private
 communication],  (which is absent in our large-$N$ approach,
 presumably due to the fact that conduction electron self-energy is
 $O(1/N)$), rather than the deconfined criticality discussed in this work.

\subsection{B. Review of the Arovas-Auerbach approach}
Here, we review the Arovas-Auerbach \cite{Arovas1988} approach to 1D antiferromagents. In the momentum space we have
\be
H=\sum_\alpha \lambda b\dg_{k\alpha} b\dn_{k\alpha}+\sum_{\alpha>0}\Delta_k(b\dg_{k\alpha}b\dg_{-k,-\alpha}+h.c.)-2S\lambda.
\ee
Here, the Lagrange multiplier enforces the constraint $n_B=2S=2sN$ necessary for a Schwinger boson representation of the spin. Using a Bogoliubov-de Gennes (BdG) transformation, we have
\be
\mat{b_{k\alpha} \\ b\dg_{-k,-\alpha}}=\mat{\cosh\theta_k & \sinh\theta_k \\ \sinh\theta_k & \cosh\theta_k}\mat{\beta_{k\alpha} \\ \gamma\dg_{k\alpha}}
\ee
We can diagonalize the Hamiltonian by tuning $\tanh2\theta_k=-\Delta_k/\lambda$ with energy $E_k=\sqrt{\lambda^2-\Delta_k^2}$, so that
\be
H=\sum_{k\alpha} [E_k(\beta\dg_{k\alpha} \beta_{k\alpha}+\gamma\dg_{k\alpha}\gamma\dn_{k\alpha})+(E_k-\lambda)]-2S\lambda
\ee
{\it Constraint} - We can write
\bea
2s&=&\frac{1}{NL}\sum_{k\alpha}\braket{b\dg_{k\alpha} b\dn_{k\alpha}}\\
&=&\frac{1}{L}\sum_k[\cosh^2\theta_k\braket{\beta\dg\beta}+\sinh^2\theta_k\braket{\gamma\gamma\dg}]\nonumber
\eea
which at zero temperature becomes
\be
4s=\frac{1}{L}\sum_k[\cosh2\theta_k-1].
\ee
In 1D we have $\Delta_k=\Delta_B\sin k$.
Since $\cosh2\theta_k=\lambda/E_k$, using $\kappa=\Delta_B/\lambda<1$
we then find
\be
4s+1=\intinf{\frac{dk}{2\pi}}\frac{1}{\sqrt{1-\kappa^2\sin^2k}}=\frac{2}{\pi}K(\kappa),
\ee
where $K(\kappa)$ is the complete elliptic integral of first kind. Parametrizing $\kappa=\cos(\vartheta)$, near $\kappa\sim1$ we have \be
4s+1=\frac{2}{\pi}K(\cos\vartheta)\approx-\frac{2}{\pi}\log(\vartheta/2),
\ee
so that the gap in the spectrum is proportional to
\be
E_g=2\Delta_B\tan(\vartheta)\approx \Delta _Be^{-\pi(1+4s)/2}.
\ee
\subsection{C. Details of large-$N$ compuations}
The self-energies of spinons and bosons are
\be 
\Sigma_B(\tau)=-2s g_c(\tau)G_\chi(\tau),\quad \Sigma_\chi(\tau)=g_c(-\tau)G_\chi(\tau).\nonumber
\ee
When written in the frequency domain and analytically continued 
onto the real frequency axis, 
the self energies become 
\bea
\Sigma_B(\omega+i\eta)&=&\gamma\intinf{\frac{d\omega'}{\pi}}f(\omega')\Big\{g_c''(\omega')G^R_\chi(\omega-\omega')\nonumber\\
&&\hspace{2cm}-g^R_c(\omega+\omega')G_\chi''(-\omega')\Big\},\qquad\label{eqSigB}\\
\Sigma_\chi(\omega+i\eta)&=&\intinf{\frac{d\omega'}{\pi}}\Big[-G_B^R(\omega+\omega')f(\omega')g_c''(\omega')\nonumber\qquad\\
&&\qquad+G_B''(\omega')n_B(\omega')g_c(\omega'-\omega-i\eta)\Big].\label{eqSigX}
\eea
In this expression all the functions are retarded ($G^{R} (\omega)\equiv G
(\omega+i\delta )= G' (\omega)+i G'' (\omega)$) and 
\be
g_c(z)= \sum_k \frac{1}{z-\epsilon_k}.\label{eqgc}
\ee
The propagator for the holon is local:
\begin{equation}
G_{\chi }^{-1} (z) = -\frac{1}{J} - \Sigma_{\chi } (z).\label{eqDysonX}
\end{equation}
By contrast, the p-wave pairing in the spinon 
Hamiltonian gives rise to an anomalous
pairing component in the spinon  Green's
functions, causing the spinons to delocalize. The $G_B(z)$ appearing in the self-energies is the local
Green's function in which the anomalous components cancel out due to
momentum-summation, \be G_B(z)\equiv \Big[{\bb G}_B^{\rm loc}(z)\Big]_{pp}.
\ee To see this explicitly, the spinon Greens function is given by \be
{\bb G}_{loc}(z)=\sum_q\frac{1}{z\tau^z-{\bb H}_B-{\bb
\Sigma}_B}.\label{eq18} \ee This momentum-sum can be performed
analytically which significantly accelerates the computation. First we
write \be {\bb G}_B(q,z)=\frac{1}{z\tau^z-{\bb H}_B-{\bb
\Sigma}_B}=\frac{1}{\Omega_B\tau^z-\Lambda_B\bb 1+\Delta_q\tau^x},\nonumber
\ee where we have used the fact that the self-energy given in 
Eq.\,\pref{eqSigB} is diagonal in the Nambu space, involving 
two components, $\bar {\Sigma }_{B}$ and
$\delta \Sigma_{B}$,
\begin{equation}\label{}
{\bb\Sigma}_{B}=\bar\Sigma_{B}\bb 1+\tau^z\delta\Sigma_{B},
\end{equation}
and we have defined
\be
\qquad \Omega_B\equiv z-\delta\Sigma_B, \qquad \Lambda_B\equiv\lambda+\bar\Sigma_B.
\ee
Using this short-hand notation, we have
\be
{\bb G}(q,z)=\frac{\Omega_B\tau^z-\Delta_q\tau^x+\Lambda_B\bb 1}{\Omega_B^2+\Delta_q^2-\Lambda_B^2}=\frac{\bb b+\bb c\sin q}{\cos 2q+a},
\ee
where we have defined
\begin{eqnarray}\label{l}
 a&=&\frac{2}{\Delta_B^2}[\Lambda_B^2-\Omega_B^2]-1,\cr 
\bb b&=&-\frac{2}{\Delta_B^2}[\Omega_B\tau^z+\Lambda_B\bb 1],
\hspace{.2cm} \bb c=-\frac{2\tau^x}{\Delta_B}. \hspace{.2cm}
\end{eqnarray}

The momentum-sum gives
\bea
{\bb G}_{loc}(z)&=&\sum_q{\bb G}(q,z)=\int_{-\pi}^{\pi}\frac{dq}{2\pi}\frac{\bb b+\bb c\sin q}{\cos 2q+a}\\
&=&\int_{-\pi}^{\pi}\frac{\bb b}{\cos q+a}=\frac{\bb b}{a\sqrt{1-a^{-2}}},\label{eq24}
\eea
which is diagonal in Nambu space with the particle-particle element proportional to  $b_{pp}={\rm Tr}[b (1+\tau_{3})]/2= -2[\Omega_B+\Lambda_B]/\Delta_B^2$.\\

{\it Saddle-point equations} - Stationarity of the Free energy with respect to $\lambda$ and $\Delta$ then 
leads to the saddle-point equation
\be
-\intinf{\frac{d\omega}{\pi}n_B(\omega)\im{G^{pp}_{B,loc}(\omega+i\eta)}}=2s,\label{eqcons}
\ee
\be
\frac{1}{2J_{H}} =
-\int{\frac{2d\omega}{\pi\Delta_B }}[n_B(\omega)+\frac{1}{2}]{\rm Im}\Big\{\sum_p\sin pG^{ph}_B(p,\omega+i\eta)\Big\},\nonumber
\ee
which determine $\lambda$ and $J_{H}$ self-consistently. The momentum sum can be done analytically and we find 
\be
\frac{1}{J_{H}} =
\int{\frac{4d\omega}{\pi\Delta_B }}[n_B(\omega)+\frac{1}{2}]{\rm Im}\Big\{\frac{1}{\Delta_B}\frac{1+a}{a\sqrt{1-a^{-2}}}\Big\}.\label{eqJH}
\ee

The calculations start at high temperature $T\gg T_K$ and the temperature is gradually reduced. Due to development of sharp features in the spectrum at low-temperatures and the importance of high-frequency components, more frequency points are required to achieve lower temperatures, a time/memory cost that eventually limits the lowest temperatures achieved.

\begin{figure}[h!]
\includegraphics[width=1\linewidth]{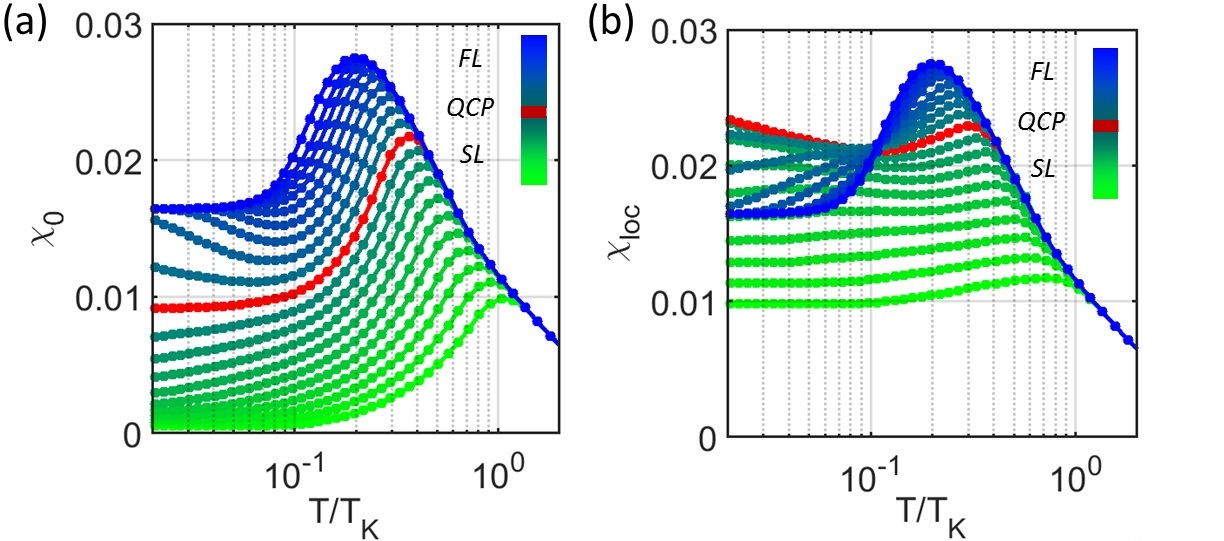}
\caption{\small (a) The uniform $\chi_0$ and (b) the local $\chi_{loc}$ spin susceptibility vs. $T/T_K$ for various values of the tuning parameter $T_K/J_H$. While  $\chi_{loc}$ diverges logarithmically at the QCP, the uniform susceptibility is only suppressed by the magnetism.}\label{figs1}
\end{figure}

Having computed the Green's functions, we can use them to calculate various thermodynamical properties:

{\it Susceptibility} - 
The spin correlation function in SP($N$) is
\bea
\chi(n,\tau)&\equiv&\sum_{\alpha\beta}\braket{-TS^{\alpha\beta}(n,\tau)S^{\beta\alpha}}\\
&=&-\tr{\bb G_B(-n,-\tau)\tau^z \bb G_B(n,\tau)\tau^z}.
\eea
The static susceptibility can be obtained by doing the imaginary-time integral of this correlation function which can be written in the form 
\bea
\chi(q,0)&=&\intob{d\tau}\sum_{\alpha\beta}\braket{-TS^{\alpha\beta}(n,\tau)S^{\beta\alpha}}\\
&=&
\int{\frac{d\omega}{\pi}}n_B(\omega)\chi''_{B}(q,\omega),\label{eq29}
\eea
where
\be
\chi_{B}(q,\omega)\equiv\sum_k\tr{{\bb G}_B(k+q,\omega+i\eta)\tau^z{\bb G}_B(k,\omega+i\eta)\tau^z}.\label{eq30}
\ee
As in the previous case, the momentum-sum can be performed analytically. For uniform and staggered susceptibility this gives
\begin{eqnarray}\label{l}
\chi_{B}(0/\pi,\omega)&=&\frac{8}{\Delta^4}\frac{\Omega_B^2+\Lambda_B^2\mp(1+a_B)\Delta_B^2/2}{a_B^2(1-a_B^{-2})^{3/2}}\cr
&\pm&\frac{4}{\Delta_B^2}\frac{1}{a_B\sqrt{1-a_B^{-2}}}.
\end{eqnarray}
For the local susceptibility, we sum Eq.\,\pref{eq30} over $q$ and find
\bea
\chi_{B}^{\rm loc}(\omega)&=&\sum_k \tr{\bb G_B^{\rm loc}(\omega+i\eta)\tau^z\bb G_B^{\rm loc}(\omega+i\eta)\tau^z},\\
&=&[G_{B,loc}^{pp}]^2+[G_{B,loc}^{hh}]^2
\eea
to be inserted in Eq.\,\pref{eq29}. Fig.\,\ref{figs1}(a,b) show the uniform and local spin susceptibilities. While $\chi_{loc}$ diverges at the QCP, $\chi_0$ is only suppressed to zero by the antiferrmagnetism.

\begin{figure}
\includegraphics[width=0.75\linewidth]{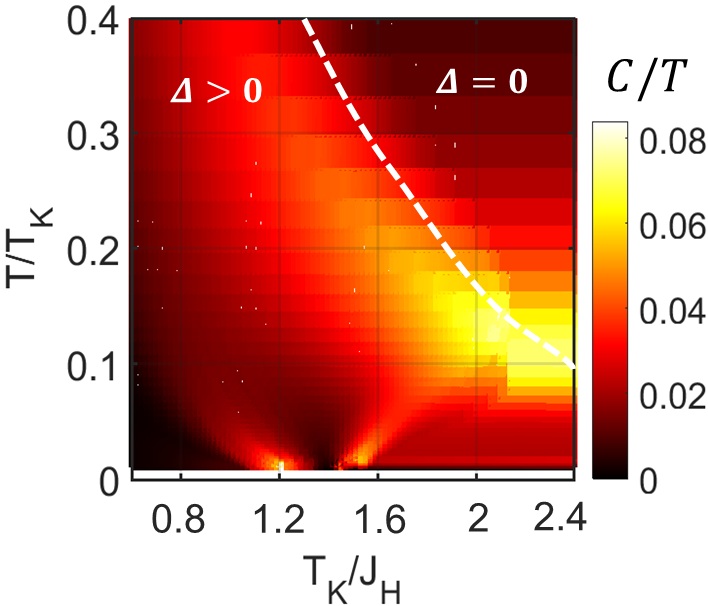}
\caption{\small The specific heat coefficient $C/T=dS/dT$ vs. $T_K/J_H$ and $T/T_J$ shows the collapse of the energy scale from both sides as well as a suppression of the $C/T$ at the QCP, consistent with a zero-point entropy.}\label{figs2}
\end{figure}

{\it Entropy} - The derivation of the entropy has been discussed in \cite{Rech2006} and \cite{Komijani18}.  The free energy can be expressed as a stationary
functional with respect to $G_B$, $G_\chi$ and $\lambda$ \cite{Komijani18}. Keeping those constant, we take derivative w.r.t $T$ to obtain the entropy. The result is \cite{Rech2006}
\bea
S (T)&=&-\int\frac{d\omega}{\pi}\partial_Tn_B(\omega)\Big\{\im{{\cal S}_{B}(\omega)}+\Sigma''_B(\omega)G'_B(\omega)\Big\}\nonumber\\
&&-k\int\frac{d\omega}{\pi}\partial_Tf(\omega)\Big\{\im{\log[-G_{\chi}^{-1}(\omega)]}\nonumber\\
&&\hspace{3cm}+\Sigma''_{\chi}(\omega)G'_{\chi}(\omega)\nonumber\\
&&\hspace{3cm}-g''_{c}(\omega)\tilde\Sigma'_c(\omega)\Big\}.\label{eq86}
\eea
Here, all the greens functions are retarded $G(\omega)\equiv G(\omega+i\eta)$ and $g_c$ is the bare Green's function of conduction band \pref{eqgc}. And
$\tilde\Sigma_c(\tau)=N\Sigma_c(\tau)$ where
$\Sigma_c(\tau)=\frac{1}{N}G_B(\tau)G_\chi(-\tau)$ is the self-energy
of the conduction electrons, which in the frequency domain is
\bea
\tilde\Sigma_C(\omega+i\eta)&=&\int{\frac{d\nu}{\pi}}\Big[n_B(\omega)G''_B(\nu+i\eta)G_\chi(\nu-\omega-i\eta)\qquad\\
&&\qquad -f(\nu)G''_\chi(\nu)G_B(\omega+\nu+i\eta)\Big].
\eea
The function ${\cal S}_{B} (z)$ is part of the spinon contribution to the entropy that requires a momentum-summation:
\bea
{\cal S}_{B} (z)&=&\frac{1}{2}\sum_q\log\Big[\det\{-G_B^{-1}(z,q)\}\Big].
\eea
Again, this momentum sum can be done analytically to obtain
\bea
{\cal S}_{B} (z)&=&\frac{1}{2}\sum_q\log[\Lambda_B^2-\Omega_B^2-\Delta_B^2\sin^2q]\\
&=&cte+\frac{1}{2}\sum_q\log[\cos 2q+a]\\
&\to &\frac{1}{2}\int{da}\sum_a\frac{1}{\cos q+a}\\\
&=&\frac{1}{2}\log[a(1+\sqrt{1-a^{-2}})].
\eea
The specific heat is obtained from numerical differentiation of the $S(T)$. Fig.\,\ref{figs2} shows the specific heat coefficient as a function of $T_K/J_H$ and $T/T_K$. The collapse of energy scale from both sides are visible. The suppression of the $C/T$ at the QCP implies a zero-point entropy due to entropy balance.

\subsection{D. The leading order solution}
Although, we have numerically obtained the solution to the self-consistent large-$N$ equations, it is interesting to study the results of a single-iteration of these equations.  

In a simple  antiferromagnet, the particle-particle component of the spinon Green's function is 
\begin{equation}
G^{pp}_B  (q,z) = \frac{z+\lambda }{z^{2}-E_q^{2}} = \frac{u_q^2}{z-E_q}-\frac{v_q^2}{z+E_q},
\end{equation}
where $E_q = \sqrt{\lambda^{2}-\Delta_q^{2}}$, $u=\cosh\theta_q$, $v=\sinh\theta_q$ and $\tanh2\theta_q=-\Delta_q/\lambda$. Substituting this into (\ref{eqSigX}), we obtain
\bea
\Sigma_{\chi }(z) = \sum_{k,q}
\Big\{&&\frac{u^2_{q}[f(\eps_k)+n_B(E_q)]}{z- (E_{q}-\epsilon_{k})} \nonumber \\
&&\hspace{1.5cm}+\frac{v_{q}([1-f(\eps_k)+n_B(E_q)]}{z- (\epsilon_{k}+E_{q})}\Big\}.\qquad
\eea
Using a flat density of states for $\eps_k$, at $T=0$ we find
\bea
\Sigma_{\chi } (z)  = - \rho \sum_{q} \Big\{&&u^2_{q}\ln
\left(
\frac{D}{E_{q }-z} \right) 
- v^2_q\ln  \left(\frac{D}{E_{q }+z}
 \right) \Big\}. \qquad
\eea

To obtain a rough estimate of this quantity, we 
can replace $E_{q}\sim \Delta $ by the typical energy of a spinon,
so that 
\bea
G_{\chi }^{-1} (z)  &=& -\frac{1}{J}+
 \rho \left[u^{2}\ln
\left(
\frac{\Lambda }{\Delta -z} \right) - v^{2}\ln  \left(\frac{\Lambda}{\Delta +z} \right) \right]
. \cr
&=& 
\rho \left[ 
u^{2}\ln  \left(\frac{T_{K}}{\Delta -z} \right) 
-v^{2}\ln  \left(\frac{T_{K}}{\Delta +z} \right) \right],
\eea
where we have used $u^{2}-v^{2}=1$, $T_{K}= \Lambda e^{-1/\rho J}$.
The phase shift at zero frequency is given by
\bea
\delta_{\chi} &=&{\rm Im}\ln [G_{\chi }^{-1} (\omega+ i\eta
)]_{\omega =0}\cr
&=& {\rm Im}\ln \left[ \ln  \left(\frac
{\Delta +i\delta }{T_{K}}
 \right) 
\right]\cr
&=& 
\pi \theta (T_{K}-\Delta ).
\eea
While this is a crude approximation, it captures the 
key feature that the holon field $\chi $ first develops a bound-state when
$T_{K}$ becomes of order $\Delta $.

\begin{figure}[t!]
\includegraphics[width=\linewidth]{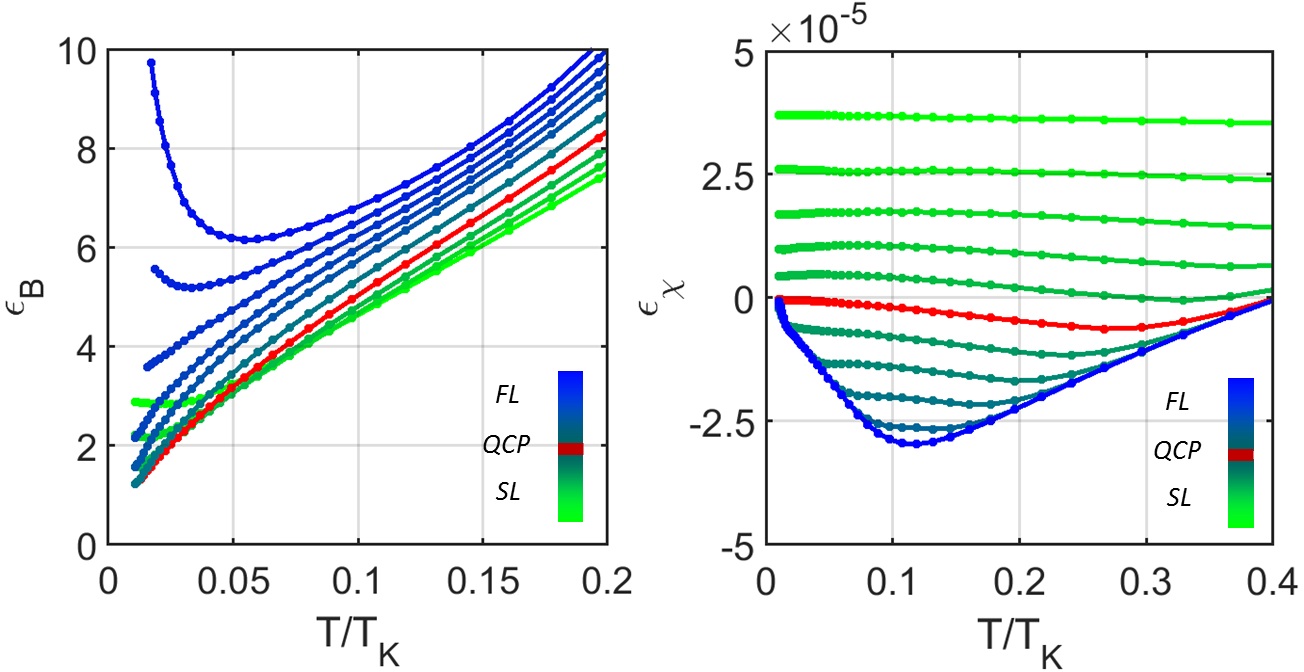}
\caption{\small (a) A plot of $\varepsilon_B\equiv \lambda+\Sigma'_B(0)-\Delta$ vs. $T$ for various values of Doniach parameter. (b) A plot of $\varepsilon_\chi\equiv 1/J+\Sigma'_\chi(0)$ vs $T$ in various regimes. }\label{figs3}
\end{figure}

\subsection{E. Scaling conditions}
Fig.\,\ref{figs3} shows the renormalized energies of spinons $\varepsilon_B\equiv \lambda+\Sigma'_B(0)-\Delta$ and holons $\varepsilon_\chi\equiv 1/J+\Sigma'_\chi(0)$. We find that $\varepsilon_B\to 0$ and $\varepsilon_\chi\to 0$ in the $T\to 0$ limit at the QCP, whereas in SL, or FL regimes, $\varepsilon_B$ remains finite.

In the FL regime, $\varepsilon_\chi$ becomes negative at a finite temperautre, indicating electron-spinon bound state formation. At the QCP, $\varepsilon_\chi\to 0^+$ as $T\to 0$ meaning that $\delta\Sigma_\chi G_\chi=-1$, whereas in the SL regime, $\varepsilon_\chi$ remains finite.\\

\begin{figure}[t!]  \includegraphics[width=\linewidth]{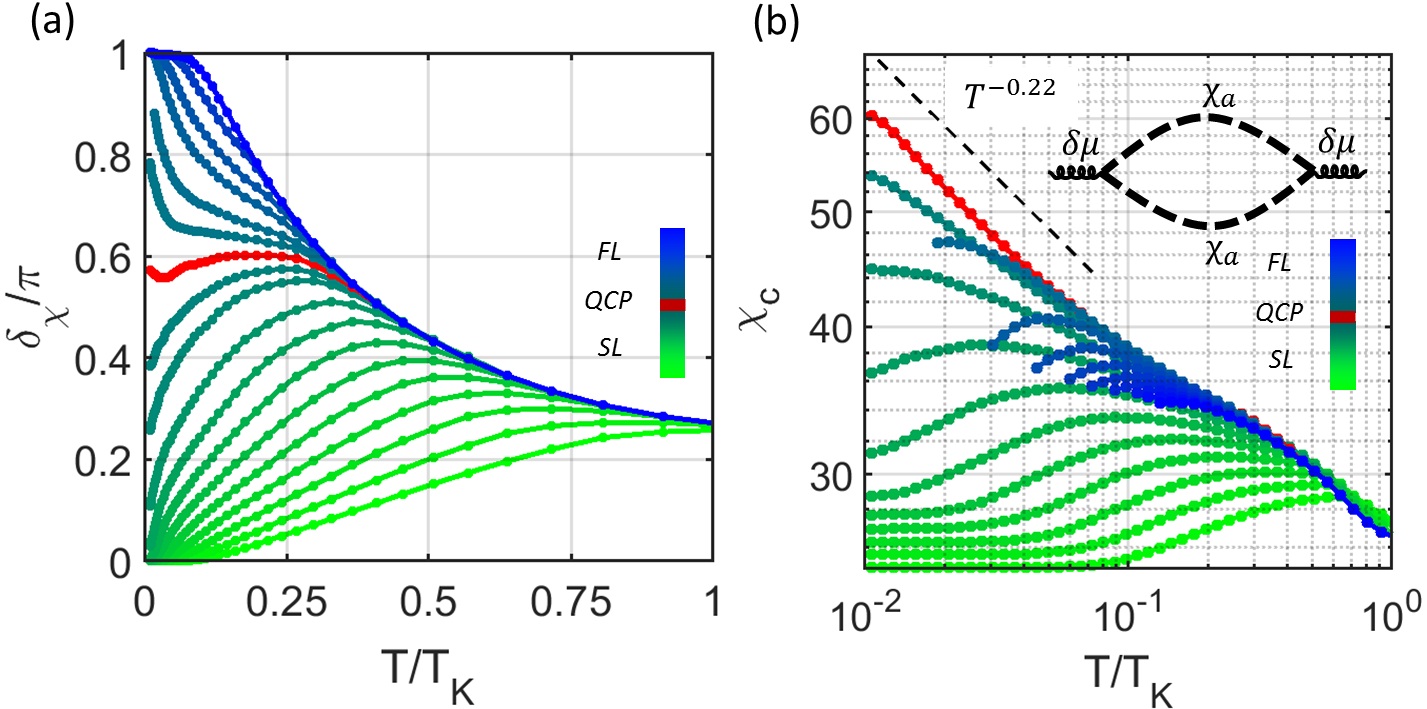}
\caption{\small (a) The holon phase shift $\delta_\chi/\pi$ as a
function of $T/T_K$, as $T_K/J_H$ is tuned from SL (green) to FL
(blue) passing SM (red). (b) The holon-bubble vs. $T/T_K$ as the
Doniach parameter is tuned from SL (green) to FL (blue) passing SM
(red). The exponents at the SL-SM transition has a $T^{-0.2}$ divergence in agreement
with the exponent extracted from $\omega/T$-scaling.}\label{figs4}
\end{figure}

The collapse of the holon spectrum on the form $G''_\chi\sim
T^{-\alpha}f(\omega/T)$, (where $\alpha=0.6$ at the specific value of
$s= S/N=0.1$ studied in the numerical work presented here) implies that at zero temperature,
$G_\chi (\omega)\sim \omega^{-\alpha }\Rightarrow G_{\chi } (\tau )\sim  \tau^{\alpha -1}$. We showed that at the QCP $\varepsilon_\chi \to 0$. Therefore, the scaling
limit of NCA equations can be used to find \bea
&G_\chi\sim \omega^{-\alpha }\sim \tau^{\alpha -1}, 
\qquad & \Sigma_B\sim\tau^{\alpha -2}\sim\omega^{1-\alpha }\\
&\Sigma_\chi\sim\omega^{\alpha }\sim\tau^{- (1+\alpha )}, 
\qquad & G_B\sim\tau^{-\alpha }\sim\omega^{\alpha -1},
\eea
at the QCP, {\it without} invoking the relation between $G_B$ and
$\Sigma_B$. The fact that $\varepsilon_B\to 0$ at the QCP and
$\omega$-powers are cancelled in $\Sigma_BG_B$ shows that the QCP is a
local one. This is in contrast to the FM case \cite{Komijani18} where
$G_B\sqrt{\Sigma_B}$ was involved. 

The leading critical exponent of $G_B\sim\tau^{-\alpha }$ contributes a
temperature-independent term to the local susceptibility
$\chi_{loc}=\intob{d\tau G_B(\tau)G_B(-\tau)}$ that depends on the UV cut-off. The log-behavior of $\chi_{\rm loc}$ shows that the sub-leading term in $G_ B$ must contain a $\tau^{-0.5}$ term. 

From $G_\chi\sim\tau^{\alpha -1}$ it follows that the holon bubble defined as $\chi_c=\intob{d\tau G_\chi(\tau)G_\chi(-\tau)}\sim T^{1-2\alpha }$ diverges at low-temperatures, which is indeed confirmed numerically as we see in the next section.

\subsection{F. Additional data on holons and spinons} Fig.\,\ref{figs4}(a) shows
the holon phase shift $\delta_\chi/\pi$ as a function of $T/T_K$ for
various values of the Doniach parameter $T_K/J_H$ tuned from FL to SL
through the strange metal phase. Note that except at the SL-SM transition, all other phase shifts
have a finite slope with temperature (at low-temperature) and approach
the two fixed points logarithmically. Fig.\,\ref{figs4}(b) shows the
holon bubble $\chi_c=\intob{d\tau G_\chi(\tau)G_\chi(-\tau)}$ as
$T_K/J_H$ is tuned between FL to SL through the QCP. The divergence at the QCP fits a $T^{-0.22}$ power-law in good agreement with the $T^{-0.2}$
power deduced from the $\omega/T$-scaling. 

A derivation of full scaling form of Green's functions (for arbitrary ratio of $\omega/T$) is beyond the scope of this work and we postpone it to the future.


\end{document}